\documentclass[letterpaper, 10 pt, conference]{ieeeconf}  

\IEEEoverridecommandlockouts                              

\usepackage{graphics} 
\usepackage{epsfig} 
\usepackage{amsmath} 
\usepackage{amssymb}  

\usepackage[usenames]{color}

\title{\LARGE \bf
Stability analysis and stabilization of systems with hyperexponential rates*
}

\author{Konstantin Zimenko$^{\dagger}$, Denis Efimov$^{\ddagger, \dagger}$, and Andrey Polyakov$^{\ddagger, \dagger}$
\thanks{*This work is supported by RSF under grant 21-71-10032 in ITMO University.}
\thanks{$^{\dagger}$Konstantin Zimenko, Denis Efimov and Andrey Polyakov are with Faculty of Control Systems and Robotics, ITMO University.}
\thanks{$^{\ddagger}$Andrey Polyakov and Denis Efimov are with 
Inria, Univ. Lille, CNRS, UMR 9189 - CRIStAL, F-59000 Lille, France.}
\thanks{\tt\footnotesize e-mail: 	
konstantin.zimenko@itmo.ru, denis.efimov@inria.fr, andrey.polyakov@inria.fr 
}}

\begin{document}

\maketitle
\thispagestyle{empty}
\pagestyle{empty}

\begin{abstract}
Hyperexponential stability is investigated for dynamical systems  with the use of both, explicit and implicit, Lyapunov function methods. A nonlinear hyperexponential control is designed for stabilizing linear systems. The tuning procedure is formalized in LMI form.  Through numeric experiments, it is observed that the proposed hyperexponential control is less sensitive with respect to noises and discretization errors than  its finite-time analog. It also demonstrates better performance in the presence of delays as well. Theoretical results are supported by numerical simulations.
\end{abstract}

\section{Introduction}

A nonlinear system is said to be fast (or hyperexponential) if its transients are faster than any linear one \cite{Caraballo, Palamarchuk, Polyakov_Fast_C}. 
Such control/observer design problems have often been the object of research, as they provide rapid error convergence (i.e., faster than any exponential)  and useful robust properties
(see, e.g., \cite{bernstein, Orlov2004, Polyakov_Fast_C, ILF_Automatica} and references therein). 
Most of the research is devoted to ensuring finite/fixed-time stability which guarantees the termination of all transients in a finite time. 

Interest in a weaker type of stability, a hyperexponential one, has increased in recent years. 
This interest is mostly related to time-delay and discrete-time systems and based on the fact that finite/fixed-time stability (stabilization) is not natural for these classes of models, and it is only possible in some rare special cases \cite{Efimov_FT_delay, Nekhoroshikh2020}, in contrast to the hyperexponential stability (see, for example, \cite{Polyakov_ILF_TD, Polyakov_CDC_17}). 

This paper investigates hyperexponential stability for systems without delays. Hyperexponential stability conditions are proposed for both, rated and unrated cases.
It is shown, that based on Implicit Lyapunov Function (ILF) method finite- and fixed-time controls (as in \cite{ILF_Automatica, mimo, Zimenko_TAC}) under sampled-time realization provide hyperexponential stability only.
In this regard, the question arises whether hyperexponential, but not finite-time, control can have better robust/discretization properties?  
The answer is yes, and it is investigated below. 

An ILF-based hyperexponential but not finite-time control is proposed for linear systems. It is recognized that proposed hyperexponential control is less sensitive with respect to noises than  its finite-time analog. Moreover, under sampled-time realization it preserves the hyperexponential stability. In that case, further development and robustness investigation of hyperexponential controls is promising, since in some cases it will allow to obtain in practise better transients and robust properties in comparison with finite-time ones.

The outline of this paper is as follows. Notation used in the paper and preliminary results on ILF method are given in Section II. Hyperexponential and non-asymptotic stability conditions are presented in Section III. Section IV presents the main results on hyperexponential stability conditions and hyperexponential control design for linear systems. Finally, concluding remarks are discussed in Section V.


\section{Preliminaries}

\subsection{Notation}

Through the paper, the following notation will be used:
\begin{itemize}
    \item $\mathbb R$ is the set of real numbers, $\mathbb R_+ =\{x\in\mathbb R : x > 0\}$;
    \item $\mathbb N$ is the set of natural numbers;
    \item the symbol $\overline{1, m}$ is used to denote a
sequence of integers $1,...,m$;
\item the order relation $P > 0$ ($< 0$; $\geq 0$; $\leq 0$) for $P \in\mathbb{R}^{n\times n}$ means that $P$ is symmetric and positive (negative) definite (semidefinite);
\item the minimal and maximal eigenvalues of a symmetric matrix $P$ are denoted by $\lambda_{\min}(P)$ and $\lambda_{\max}(P)$, respectively;
\item $\text{diag} \{\lambda_i\}_{i=1}^n$ is a diagonal matrix with elements $\lambda_i$.
\end{itemize}

\subsection{Implicit Lyapunov function method}

Consider the system of the form
\begin{equation}
\label{sys}
    \dot x(t) =f(x(t)), \quad x(0)=x_0, \quad t\geq 0,
\end{equation}
where $x\in\mathbb R^n$ is the state vector, $f:\mathbb R^n\to \mathbb R^n$ is a nonlinear vector field, $f(0)=0$, which can be discontinuous with respect to the state variable. In this case the solutions $\Phi(t, x_0)$ of the system~(\ref{sys})
 are understood in the sense of Filippov \cite{Filippov}.

The next theorem presents the Implicit Lyapunov function method \cite{Korobov, Adamy_Flemming}.

\textbf{Theorem 1} \textit{Let there exists continuous function $Q\colon \mathbb{R}_{+} \times \mathbb{R}^n \to \mathbb{R}$ satisfying the conditions\\
\textit{C1)} $Q$ is continuously differentiable outside the origin of $\mathbb{R}_{+}\times \mathbb{R}^n \backslash\{0\}$;\\
\textit{C2)} for any $x \in \mathbb{R}^n \backslash\{0\}$ there exists $V\in \mathbb{R}_{+}$ such that $Q(V,x)=0$;\\
\textit{C3)} for $\Omega=\{(V, x) \in\mathbb R_+\times \mathbb{R}^{n} \colon Q(V, x)=0\}$
$$
\lim_{\substack{x\to 0 \\ (V, x) \in \Omega}}V=0^{+}, \lim_{\substack{V\to 0^{+} \\ (V, x) \in \Omega}}\|x\|=  0, \lim_{\substack{\|x\| \to \infty \\ (V, x) \in \Omega}}V  = +\infty;
$$
\textit{C4)} the inequality 
$\frac{\partial Q(V, x)}{\partial V}<0$
holds for all $ V \in \mathbb{R}_{+}$ and $\forall x \in \mathbb{R}^n \backslash\{0\}$.\\
If $\frac{\partial Q(V, x)}{\partial x}f(x) < 0$
for all $(V, x) \in \Omega$, then the origin of~(\ref{sys}) is globally 
asymptotically stable.
}


\section{Hyperexponential and non-asymptotic stability definitions}

 Over the past several decades, increased interest has been paid to stability notions with non-asymptotic convergence, e.g., finite-time and fixed-time stability (see, \cite{bernstein, Orlov2004, ILF_Automatica, fixattr, Zimenko_TAC} and references therein).
 
  Let $D$ be an open neighborhood of the origin in $\mathbb R^n$.
 
 \textbf{Definition 1 \cite{bernstein, Orlov2004}} \textit{The origin of~(\ref{sys}) is said to be   finite-time stable if it is  asymptotically stable and for any $x_0\in D$  any solution $\Phi(t, x_0)$ of the system~(\ref{sys}) reaches the origin at some finite time moment, i.e. $\Phi(t,x_0)=0 \; \forall t\ge T(x_0)$ and $\Phi(t, x_0)\neq 0 \; \forall t\in [0,T(x_0))$, $x_0\neq 0$, where $T\colon \mathbb R^n \to \mathbb R_+\cup \{0\}$, $T(0)=0$ is a settling-time function.
 }
 
 \textbf{Definition 2 \cite{fixattr}} \textit{The origin of~(\ref{sys}) is said to be fixed-time stable if it is finite-time stable and the settling-time function $T(x_0)$ is bounded, i.e., $\exists T_{\max}>0 \colon T(x_0)\le T_{\max}, \; \forall x_0 \in D$.
 }
 
If $D=\mathbb R^n$, then the corresponding stability properties are called global.

 \textbf{Definition 3 \cite{Polyakov_Fast_C}} \textit{A set $\mathcal{M}\subset\mathbb R^n$ is said to be globally fixed-time attractive if any solution $\Phi(t, x_0)$ of~(\ref{sys}) reaches $\mathcal{M}$ in some finite time moment $t=T(x_0)<T_{\max}$, $T_{\max}\in\mathbb R_+$ and remains there for all $ t\ge T(x_0)$, where $T\colon \mathbb{R}^n\to\mathbb{R}_+\cup\{0\}$.  The origin is said to be nearly fixed-time stable if it is globally Lyapunov stable and any neighborhood of the origin is  fixed-time attractive.}

The interest in finite/fixed-time stability  is due to existence of  many fast control applications with time constraints, where a high precision is needed in the presence of significant perturbations (e.g., rejection of non-Lipschitz disturbances).

An extension of Theorem 1 for finite/fixed-time stability
analysis is proposed in \cite{ILF_Automatica}.

Another type of stability, which also applies to fast systems, is hyperexponential stability (see, e.g., the papers \cite{Caraballo, Palamarchuk, Polyakov_Fast_C}).

 \textbf{Definition 4 \cite{Polyakov_Fast_C}} \textit{
The system~(\ref{sys}) is said to be
     hyperexponentially stable at the origin in $D$ if the system is Lyapunov stable and
\begin{equation}
    \label{hed}
      \begin{array}{cc}
  \forall r\in\mathbb R_+\; \exists t'\in\mathbb R_+, \kappa>0,\;C>0:\\ \|\Phi(t,x_0)\|\leq Ce^{-rt},\;\forall t>t',\;\forall x_0\in\{x\in D:\|x\|\leq\kappa\}
     \end{array}
\end{equation}
  }

In order to  provide quantitative index to characterize  hyperexponential convergence rates, the rated hyperexponential stability is presented in \cite{Polyakov_ILF_TD}.

For $\alpha=(\alpha_0, \alpha_1, \cdots, \alpha_r)^\top\in\mathbb R_+^{r+1}$ and $r\in\mathbb N$
let us introduce the function of nested  exponentials $\rho_{r,\alpha}:\mathbb R\to\mathbb R_+$ by the following recursive formula
\begin{equation}
    \label{rhodef}
   \begin{array}{ll}
\rho_{0, \alpha}(z)=\alpha_0 z, \\ \rho_{i,\alpha}(z)=\alpha_i \left( e^{\rho_{i-1,\alpha}(z)}-e^{\rho_{i-1,\alpha}(0)}\right), \quad i=\overline{1,r}.
\end{array} 
\end{equation}

 \textbf{Definition 5 \cite{Polyakov_ILF_TD}} \textit{
The origin of the system~(\ref{sys}) is said to be hyperexponentially stable of degree $r\in\mathbb N$ in $D$, if it is hyperexponentially stable and for each $\kappa>0$ $\exists C\in\mathbb R_+$ and $\alpha\in\mathbb R_+^{r+1}$ such that 
\begin{equation}
    \label{hesd}
 \|\Phi(t, x_0)\|\leq Ce^{-\rho_{r,\alpha}(t)}\quad \;\forall x_0\in\{x\in D:\|x\|\leq\kappa\}. 
 \end{equation}}

Notice that
\begin{itemize}
    \item for the case $r=0$  the expression~(\ref{hesd}) corresponds to exponential stability definition;
    \item all finite-time and fixed-time stable systems are also hyperexponentially stable due to finite/fixed-time stability is obtained through an "infinite eigenvalue assignation" for the system at the origin.
\end{itemize}

  Fig. 1 depicts in a logarithmic scale $e^{-\rho_{i,\alpha}(t)}$ for $t\geq0$, $\alpha_i$, $i=\overline{0,2}$ with $\alpha = (1, 1, 1)^\top$ and $\left\{\begin{array}{ll}
(1-0.8t)^{1.25} \;\; &\text{for} \; t\in[0, 1.25],  \\
0      &\text{for} \; t>1.25\\ 
 \end{array}\right.$ to represent the finite-time stability case   in order to show the decay rates.
 
 \begin{figure}
\begin{center}
\includegraphics[width=8.5cm]{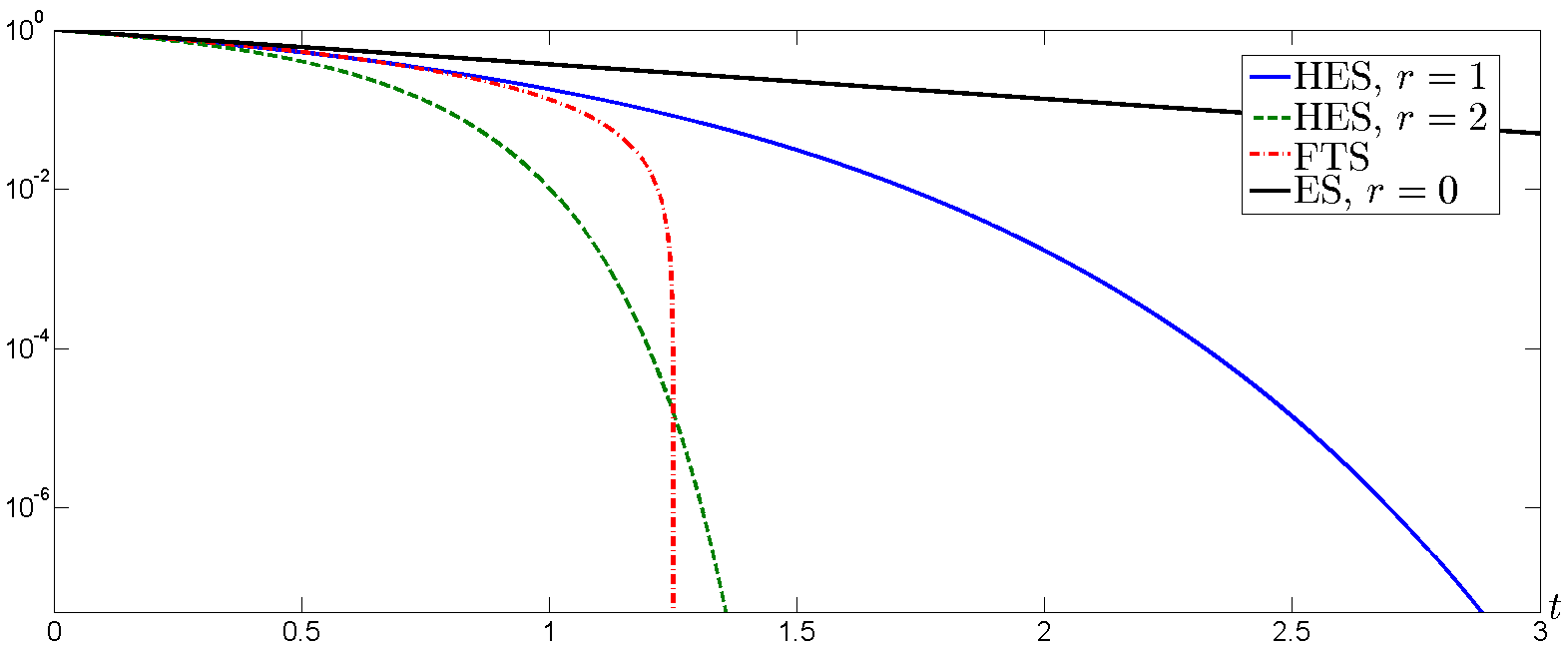}   
\caption{Rates of convergence: exponential (ES); hyperexponential (HES); finite-time (FTS)}  
\label{Fig1}                              
\end{center}                                
\end{figure}


\section{Hyperexponential stability analysis using Lyapunov function method}

Lyapunov function method \cite{Lyapunov_1992} is the main tool for stability analysis of nonlinear systems. The following theorem presentst a sufficient condition for hyperexponential stability.

\textbf{Theorem 2} \textit{
Suppose there exist $C^1$ functions $V_i$, $i=\overline{1,+\infty}$ defined on $D$ and corresponding sequence of open sets $\{D_i\}_{i=1}^{\infty}: D= D_1\supset D_2\supset ...\supset D_i\supset...\ni\{0\}$ such that $D_i=\{x\in\mathbb R^n: V_i<v_i, \; v_i\in\mathbb R_+\}$ and 
\begin{equation}
    \label{heL1}
    \dot V_{i}(x) \le -c_i V_i(x), \quad \forall x \in D_i\setminus D_{i+1}, 
\end{equation}
where $c_{i+1}>c_i\in\mathbb R_+$ for any $i\in\mathbb N$ and $c_i\to+\infty$ as $i\to+\infty$.
Then the origin of system~(\ref{sys}) is hyperexponentially stable. If $D = \mathbb R^n$ and function $V$ is radially unbounded, then the system~(\ref{sys}) admits this property globally.
}

\begin{proof}
The proof is straightforward due to the condition~(\ref{heL1}) implies that for any $i\in\mathbb N$ the set $D_i$ is strictly positively invariant and the solution $\Phi(t,x_0)$ reaches the nested set $D_{i+1}$ in a finite time with corresponding exponential rates; the condition  $c_i\to+\infty$ as $i\to+\infty$ implies increasing of the rates as solution reaches the nested set and $D_i\to\{0\}$ as $i\to\infty$ (i.e. the origin is attractive).
\end{proof}


\subsection{ILF-based finite-time control under sampled-time realization}

The following application shows that sampled-time realization of ILF-based finite-time control implies hyperexponential stability.

\textbf{Example 1} 
Consider the system 
\begin{equation}
  \label{mimosys}
  \dot x=Ax+Bu,
\end{equation}
where $x\in\mathbb R^n$ is a state, $u\in \mathbb R$ is a control input,
\begin{equation}
  \label{mimosys2}
\begin{array}{cc}
A=\begin{bmatrix} 0 & 1 & 0&\cdots &0\\
0 & 0& 1 & \cdots &0\\
\cdots & \cdots & \cdots&\cdots &\cdots\\
0 & \cdots& \cdots &0& 1\\
0 & \cdots& \cdots & 0 &0\\
\end{bmatrix}\in\mathbb R^{n\times n}, \\
B=\begin{bmatrix}0&0&\cdots&0&1 \end{bmatrix}^\top \in\mathbb R^{n\times 1}.
\end{array}
\end{equation}
Introduce the ILF function
$$
Q(V,x)=x^TD(V^{-1})PD(V^{-1})x-1,
$$
where $V\in\mathbb R_+$, 
$$
\begin{array}{cc}
D(\lambda)=\begin{bmatrix}
\lambda^{q_1} & 0 & \cdots & 0\\
0 & \lambda^{q_2} & \cdots & 0\\
\cdots & \cdots & \cdots & \cdots\\
0 & 0 & 0 & \lambda^{q_n} \\
\end{bmatrix}, \quad \lambda\in\mathbb R_+,\\
q_i=1+(n-i)\mu, \quad i=\overline{1,n},
\end{array}
$$
$\mu\in(0,1]$ and $0<P\in\mathbb R^{n\times n}$. Denote $H=\text{diag}\{q_i\}_{i=1}^n$.

According to \cite{ILF_Automatica} the  control 
\begin{equation}
    \label{contrmimo}
 u(V,x)=V^{1-\mu}KD(V^{-1})x   
\end{equation}
stabilizes the system~(\ref{mimosys}) in a finite time, where $V\in\mathbb R_+: Q(V,x)=0$ and $K=YX^{-1}$, where $X=P^{-1}\in\mathbb R^{n\times n}$, $Y\in\mathbb R^{1\times n}$ satisfy the system
$$
\begin{array}{cc}
AX+XA^T+BY+Y^TB^T+aX\leq 0,\\
XH+HX>0, \quad X>0,
\end{array}
$$
for some $a\in\mathbb R_+$.

Denote an arbitrary sequence of time instances $\{t_i\}_{i=1}^{+\infty}$ such that $0 = t_0 < t_1 < t_2 <...$, $\lim_{i\to+\infty} t_i=+\infty$. In \cite[Corollary 8]{ILF_Automatica} it is shown that sampled-time realization of the control~(\ref{contrmimo}) (i.e. $u=u(V_i,x)$ on each time interval $[t_i, t_{i+1})$, where $V_i\in\mathbb R_+:Q(V_i,x(t_i))=0$) provides asymptotic
stabilization of the closed loop system independently on the length of the sampling interval. Let us show that sampled-time realization implies \textit{hyperexponential stability} of the closed loop system.

Let us define the set of nested ellipsoids $D_i:=\{x\in\mathbb R^n: x^\top P_i x\leq 1\}$, where $P_i=D(V_i^{-1})PD(V_i^{-1})$ and the corresponding set of quadratic functions $\tilde V_i=x^\top P_ix$. Then for any $i\in\mathbb N$ on $D_i$ we have
\small
$$
\begin{array}{ll}
\dot{\tilde V}_i=2x^\top P_i(Ax+Bu(V_i,x))\\
=\!\tilde V_i^{-\mu}x^\top D(V_i^{-1})\left(PA\!+\!A^\top P\!+\!PBK\!+\!K^\top B^\top P \right)D(V_i^{-1})x\\
\leq -a\tilde V_i^{1-\mu}.
\end{array}
$$
\normalsize
Due to $V_{i}>V_{i+1}$ and $V_i\to 0$ as $i\to\infty$ (see \cite{ILF_Automatica} for more details), then applying Theorem 2 with $c_i=a\tilde V_i^{-\mu}$, the closed loop system is hyperexponentially stable. The results of simulation  are shown in  Fig. 2, Fig. 3 for $n=3$, $\mu=0.5$,
$$
P=\begin{bmatrix} 3.3119  &  3.2943  &  0.8806\\
    3.2943  &  5.2366  &  1.4426\\
    0.8806  &  1.4426  &  0.9682
\end{bmatrix},\quad K^T=\begin{bmatrix} -4.2618 \\  -7.7818 \\  -3.2635
\end{bmatrix}
$$ and sampling period $t_{i+1}-t_i =1$.
Fig. 3 shows the Lyapunov function $V(t)$ with using the logarithmic scale in order to demonstrate hyperexponential convergence rate. 

 \begin{figure}
\begin{center}
\includegraphics[width=8.5cm]{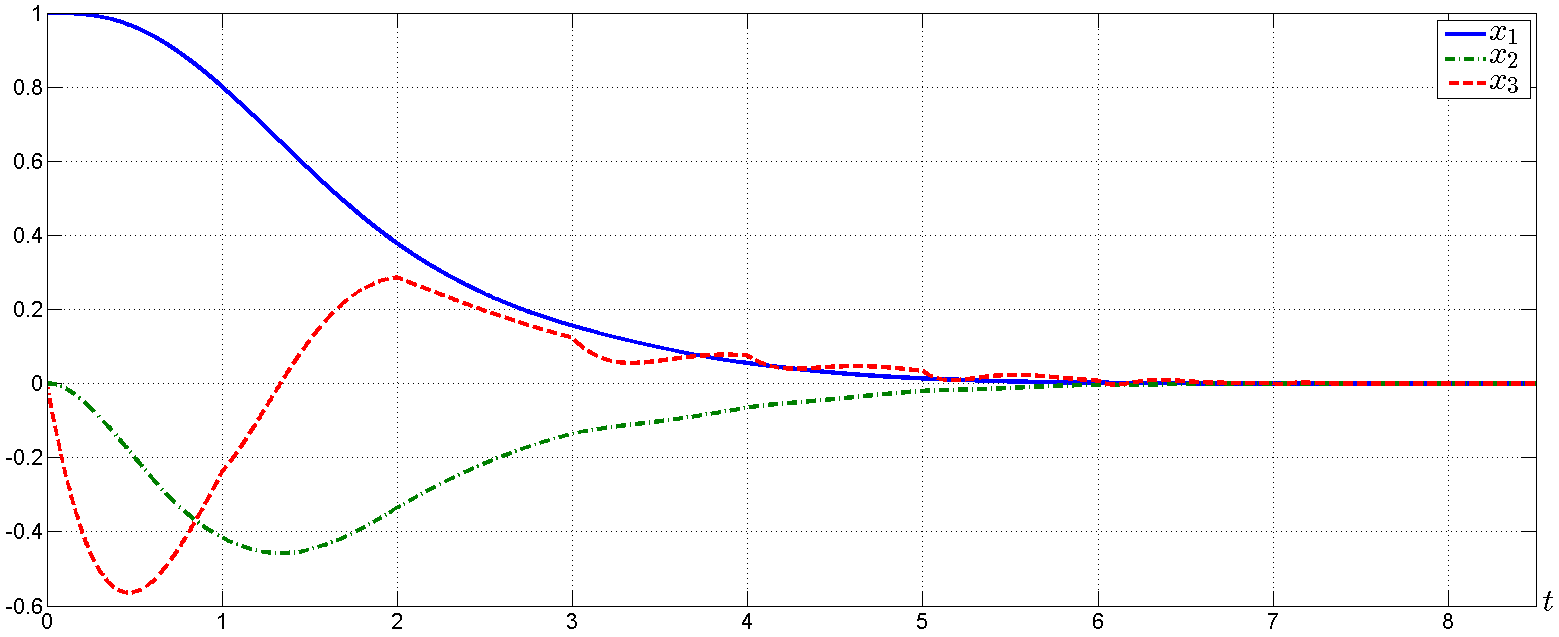}   
\caption{State transients for the system~(\ref{mimosys}) with sampled-data realization of the control (\ref{contrmimo}) }
\label{Fig2}                              
\end{center}                                
\end{figure}

 \begin{figure}
\begin{center}
\includegraphics[width=8.5cm]{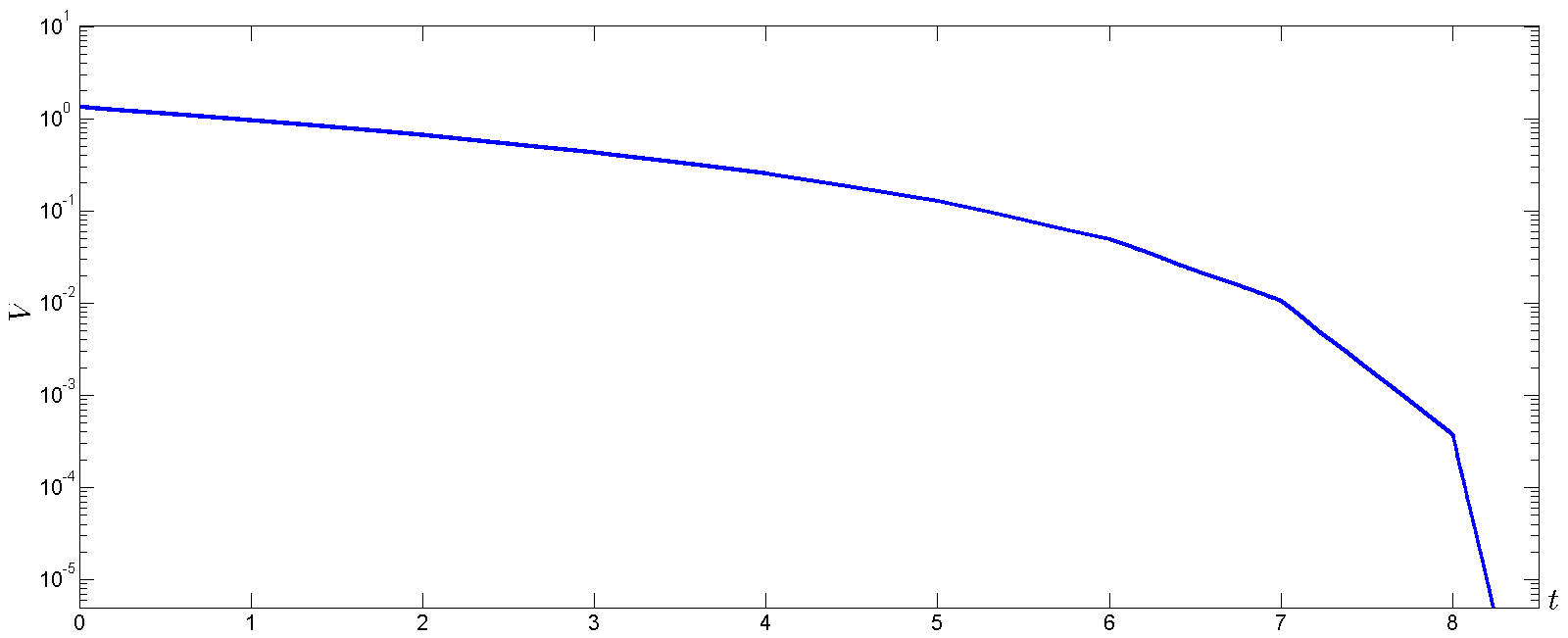}   
\caption{Lyapunov function $V(t)$ for the system~(\ref{mimosys}) with sampled-data realization of the control (\ref{contrmimo}) }
\label{Fig3}                              
\end{center}                                
\end{figure}

The same outcomes can be obtained for finite/fixed-time controls proposed in \cite{ILF_Automatica, mimo, Zimenko_TAC}. Consider the system~(\ref{mimosys}),
where 
the pair $(A,B)$ is controllable.

\textbf{Proposition 1} \textit{Let $\{t_i\}_{i=0}^\infty$ be a strictly increasing sequence of arbitrary time instants, $0 = t_0 < t_1 < t_2 <...$ such that $\lim_{i\to\infty} t_i = +\infty$. Let for~(\ref{mimosys}) all conditions proposed in \cite{ILF_Automatica} (or in \cite{mimo, Zimenko_TAC}) for ILF-based finite/fixed-time control $u(V,x)$ design  be satisfied. Then sampled-time realization $u(t) = u(V_i, x(t))$ for $t\in [t_i, t_{i+1})$, where $V_i\in\mathbb R_+ : Q(V_i, x(t_i)) = 0$ provides hyperexponential stability of the closed-loop system.}

\begin{proof}
The proof follows exactly the same arguments as given in Example 1 and ILF properties.
\end{proof}

\subsection{Explicit Lyapunov function method}

Since the hyperexponential stability assumes increasing of exponential rates as $\Phi(t,x_0)\to 0$, then the following result can be obtained:

\textbf{Theorem 3} \textit{Let $\beta:\mathbb R_+\to\mathbb R_+$ be positive nondecreasing function  
such that $\beta(s)\to\infty$ as $s\to\infty$.
Suppose there exists a positive definite $C^1$ function $V$ defined on an open neighborhood of the origin $D\subset \mathbb R^n$, such that the following condition is true for the system~(\ref{sys})
$$ \dot V(x) \le -\beta(V^{-1}(x)) V(x), \quad x \in D\setminus \{0\}. $$
Then the origin of system~(\ref{sys}) is globally hyperexponentially stable. If $D = \mathbb R^n$ and function $V$ is radially unbounded, then the system~(\ref{sys}) admits this property globally.}

\begin{proof}
Due to the function $\beta$ is positive and nondecreasing we have that $\dot V(x)\leq -C_0V(x)$, $C_0=\beta(V(x_0)^{-1})$ for all $x \in D\setminus \{0\}$, i.e. the system~(\ref{sys}) is exponentially stable. Then, due to $\beta(s)\to\infty$ as $s\to\infty$, there exists a sequence of time instances $\{t_i\}_{i=0}^{+\infty}$: 
$0 = t_0 < t_1 < t_2 <...$, $\lim_{i\to+\infty} t_i=+\infty$ such that
$$\beta(V(x(0))^{-1})\!<\!\beta(V(x(t_1))^{-1})\!<\!\cdots\!<\!\beta(V(x(t_i))^{-1})\!<\!\cdots$$
and 
$$
\dot V(x(t))\leq -C_iV(x(t)), \; C_i=\beta(V(x(t_i))^{-1}), \quad \forall t\geq t_i,
$$ 
that implies hyperexponential stability since $C_i\to+\infty$ as $t_i\to \infty$.
\end{proof}


\subsection{ILF method}

Let us introduce the functions $\sigma_i^\alpha :(0,1]\to \mathbb R$, $i=\overline{1,r}$ by the following recursive formula 
$$
\begin{array}{ll}
\sigma_1^\alpha(s)=-\ln s+\alpha_r,\\
\sigma_i^\alpha(s)=\ln\left(\frac{1}{\alpha_{r-i+2}}\sigma_{i-1}^\alpha (s) \right)+\alpha_{r-i+1}, \quad i=\overline{2,r}.
\end{array}
$$
Obviously $\sigma_i^\alpha(1)=\alpha_{r-i+1}$. 
Note that $\sigma_i^\alpha(s)>0$ for $s\leq 1$ and $\sigma_i^\alpha(s)\to+\infty$ as $s\to 0$.

The next theorem presents the ILF method for \textit{rated hyperexponential stability}.

\textbf{Theorem 4} \textit{Let there exist two functions $Q_1$ and $Q_2$ satisfying the conditions $C1)$-$C4)$ of Theorem 1 and\\
\textit{C5)} $Q_1(1,x) = Q_2(1,x)$ for all $x \in \mathbb R^n\setminus \{0\}$;\\
\textit{C6)} there exist $\alpha=(\alpha_0, \alpha_1, \cdots, \alpha_r)^\top\in\mathbb R_+^{r+1}$, $r\in\mathbb N$, $c_1\in\mathbb R_+$, 
such that the inequality 
$$
\frac{\partial Q_1(V, x)}{\partial x}f(x)\leq  c_1 V\prod_{i=1}^r \sigma_i^\alpha(V)  \frac{\partial Q_1(V, x)}{\partial V},
$$
holds for all $V\in(0, 1]$ and $x \in \mathbb R^n\setminus \{0\}$ satisfying the equation $Q_1(V,x) = 0$;\\
\textit{C7)} there exists $c_2\in\mathbb R_+$ such that the inequality 
$$
\frac{\partial Q_2(V, x)}{\partial x}f(x)\leq c_2 V  \frac{\partial Q_2(V, x)}{\partial V},
$$
holds for all $V\geq 1$ and $x \in \mathbb R^n\setminus \{0\}$ satisfying the equation $Q_2(V,x) = 0$;\\
\textit{C8)} for some $k_1, k_2, a\in\mathbb R_+$ the inequality
$$
\begin{array}{cc}
k_1 \|x\|^a\leq V\leq k_2 \|x\|^a, 
\end{array}
$$
holds for all $V\in\mathbb R_+$ and $x \in \mathbb R^n\setminus \{0\}$ satisfying the equation $Q_1(V,x) = 0$ for $V\leq 1$ and $Q_2(V,x) = 0$ for $V> 1$,
\\
then the origin of the system~(\ref{sys}) is globally hyperexponentially stable with degree
$r$.
}

\pagebreak
\begin{proof}
Let the two functions $V_1$ and $V_2$ be defined by the equations $Q_1(V_1,x) = 0$ and
$Q_2(V_2,x)=0$. Consider the sets $\Sigma_1 =\{x\in\mathbb R^n :V_1(x)>1\}$, $\Sigma_2=\{x \in\mathbb R^n : V_2(x) > 1\}$ and prove that $\Sigma_1 =\Sigma_2$. Suppose the contrary, i.e. $\exists z\in\mathbb R^n$
such that $z\in\Sigma_1$ and $z\notin \Sigma_2$. On the one hand, $Q_1(V_1, z) = 0$ implies $V_1 > 1$ and
$Q_1(1, z) > Q_1(V_1, z) = 0$ due to condition \textit{C4)}. On the other hand, $Q_2(V_2, z) = 0$
implies $V_2\leq 1$ and $Q_2(1, z)\leq Q_2(V_2, z) = 0$. The contradiction follows from \textit{C5)}.

Therefore, due to \textit{C5)} and \textit{C4)} the function $V :\mathbb R^n\to\mathbb R$ defined by the equality
$$
V(x)=\left\{\begin{array}{ll}
V_1(x)     & \;\; \text{for} \quad V_1(x)<1,\\
V_2(x)     & \;\; \text{for} \quad V_2(x)>1, \\
1 & \;\;  \text{for} \quad V_1(x)=V_2(x)=1,
\end{array} \right.
$$
is positive definite, continuous in $\mathbb R^n$ and continuously differentiable for $x\notin \{0\}\cup \{x\in\mathbb R^n: V(x)=1\}$. The function $V$ is Lipschitz continuous outside the origin and has the following Clarke’s gradient \cite{Clarke_1990}:
$$
\nabla_C V(x)=\xi \nabla V_1(x)+(1-\xi)\nabla V_2(x), \quad x\in\mathbb R^n,
$$
where $\xi = 1$ for $0<V_1(x)< 1$, $\xi = 0$ for $V_2(x) > 1$, $\xi = [0,1]$ for $V_1(x) =V_2(x) = 1$
and $\nabla V_i$ is the gradient of the function $V_i$, $i = 1, 2$. Hence, due to conditions \textit{C6)} and \textit{C7)}, the inequality
\begin{equation}
    \label{dotV}
  \frac{\partial V(\Phi(t, x_0))}{\partial t}\leq \left\{\begin{array}{ll}
 -c_1 V(\Phi(t, x_0))\prod_{i=1}^r \sigma_i^\alpha(V(\Phi(t, x_0)))  \\     \qquad \qquad\qquad \text{for} \quad V(\Phi(t, x_0))<1,\\
  -c_2 V(\Phi(t, x_0)) \\ 
  \qquad\qquad\qquad   \text{for} \quad V(\Phi(t, x_0))>1,\\
     -\min\{c_1 \prod_{i=1}^r\alpha_i, c_2\} \\
     \qquad\qquad\qquad \text{for} \quad V(\Phi(t, x_0))=1\\
\end{array} \right.  
\end{equation}
holds for almost all $t$ such that $\Phi(t, x_0)\neq 0$. 

Let us consider the scalar differential equation
\begin{equation}
  \label{comlem}
  \dot y(t)=-\alpha_0 y(t)\prod_{i=1}^r\sigma_i^{\alpha}\left(\frac{y(t)}{y_0}\right), \quad y(0)=y_0\in\mathbb R_+.  
\end{equation}
The solution of~(\ref{comlem}) is $y(t)=y_0 e^{-\rho_{r,\alpha}(t)}$. Indeed, due to 
$$
\begin{array}{ll}
 \frac{\partial \rho_{i,\alpha}(t)}{\partial t}&=\alpha_i e^{\rho_{i-1, \alpha}(t)}\frac{\partial \rho_{i-1,\alpha}(t)}{\partial t}\\
&=(\rho_{i,\alpha}(t)+\alpha_i e^{\rho_{i-1,\alpha}(0)}) \frac{\partial \rho_{i-1,\alpha}(t)}{\partial t} \\
&=(\rho_{i,\alpha}(t)+\alpha_i) \frac{\partial \rho_{i-1,\alpha}(t)}{\partial t},\\
\frac{\partial \rho_{0,\alpha}(t)}{\partial t}&=\alpha_0,  \\
\rho_{r,\alpha}(t)&=\ln\left(\frac{y_0}{y(t)} \right),\\
\rho_{i-1,\alpha}(t)&=\ln\left(\frac{1}{\alpha_i}\rho_{i,\alpha}(t)+1 \right),
\end{array}
$$
we have 
\begin{equation}
    \label{Efque}
 \begin{array}{ll}
\frac{\partial y_0 e^{-\rho_{r,\alpha}(t)}}{\partial t}&=-y(t) \frac{\partial \rho_{r,\alpha}(t)}{\partial t}\\
&= -y(t) \sigma_1^\alpha\left(\frac{y(t)}{y_0} \right) \frac{\partial \rho_{r-1,\alpha}(t)}{\partial t}\\
&= -y(t) \sigma_1^\alpha\left(\frac{y(t)}{y_0} \right)
\! \sigma_2^\alpha\left(\frac{y(t)}{y_0} \right) \frac{\partial \rho_{r-2,\alpha}(t)}{\partial t}\\
&=\cdots= -\alpha_0 y(t) \prod_{i=1}^r \sigma_i^\alpha\left(\frac{y(t)}{y_0} \right).\\
\end{array}   
\end{equation}
Thus,
the system~(\ref{comlem}) is globally hyperexponentially stable with degree
$r$ and convergence rate $\alpha$. 
Returning to~(\ref{Efque}) we yeild
$$
 \begin{array}{ll}
\frac{\partial y(t)}{\partial t}&=-\alpha_0 y(t) \prod_{i=1}^r \sigma_i^\alpha\left(\frac{y(t)}{y_0} \right)\\
&= -\alpha_0 y(t) \prod_{i=1}^r \sigma_i^\alpha\left(e^{-\rho_{r,\alpha}(t)} \right)\\
&<0 \qquad\quad \forall t\geq 0,
\end{array} 
$$
i.e., $y(t)$ is strictly decreasing function of $t$ and $y(t)\leq y_0$.
 Taking into account that $\prod_{i=1}^r\sigma_i^{\alpha}\left(\frac{y(t)}{y_0}\right)\leq \prod_{i=1}^r\sigma_i^{\alpha}\left(\frac{1}{y_0}\right)$ for $y_0\geq y(t)\geq 1$ and
$\prod_{i=1}^r\sigma_i^{\alpha}\left(\frac{y(t)}{y_0}\right)\leq \prod_{i=1}^r\sigma_i^{\alpha}\left(y(t)\right)$ for $1\geq y_0\geq y(t)$,
then by the comparison lemma we have that the system in the form
$$
\frac{\partial y(t)}{\partial t}\leq \left\{\begin{array}{ll}
 -\alpha_0 y(t) \prod_{i=1}^r \sigma_i^\alpha\left(y(t) \right)       \quad  \text{for} \quad y(t)<1,\\
 -\alpha_0 y(t) \prod_{i=1}^r \sigma_i^\alpha\left(\frac{1}{y_0} \right)  \quad\;   \text{for} \quad y(t)>1,\\
     -\min\left\{\prod_{i=0}^r\alpha_i,\; \alpha_0  \prod_{i=1}^r \sigma_i^\alpha\left(\frac{1}{y_0} \right)\right\} \\
     \qquad\qquad\qquad \text{for} \quad y(t)=1\\
\end{array} \right.
$$
is hyperexponentially stable with degree $r$ and convergence rate $\alpha$. Then, returning to~(\ref{dotV}), we have that $V(\Phi(t,x_0))$ hyperexponentially converges to $0$ with degree $r$ and  rate $\alpha=\left(c_1, \alpha_1, \alpha_2,...,\alpha_r \right)^\top$ if $c_2\geq c_1\prod_{i=1}^r \sigma_i^\alpha\left(\frac{1}{V(x_0)}\right)$ and
$ \alpha=\left(\frac{c_2}{\prod_{i=1}^r \sigma_i^\alpha\left(\frac{1}{V(x_0)} \right)}, \alpha_1, \alpha_2,...,\alpha_r \right)^\top$ if $c_2< c_1\prod_{i=1}^r \sigma_i^\alpha\left(\frac{1}{V(x_0)}\right)$, i.e., $V(\Phi(t, x_0))\leq V(x_0) e^{-\rho_{r, \alpha}(t)}$.  
Finally, according to $C8)$ we have $\|\Phi(t, x_0)\| \leq \left(\frac{k_1}{k_2}\right)^{1/a} \|x_0\| e^{-\rho_{r, \alpha}(t)/a}$.
\end{proof}

Note that due to the condition \textit{C3)} the function $V$ is radially unbounded and it can be continuously prolonged at the origin by $V(0) = 0$.

The condition $C8)$ in Theorem 4 may be relaxed as it is shown in the following corollary.

\textbf{Corollary 3} \textit{Let there exist two functions $Q_1$ and $Q_2$ that satisfy the conditions $C1)$-$C4)$ of Theorem 1, the conditions $C5)$-$C7)$ of Theorem 4 with $ r\geq 2$ and
\\
\textit{C9)} for some $k\in\mathbb R_+$ the inequality
\begin{equation}
   \label{strus}
\frac{\alpha_r}{-\ln \left(V \right)+1} \geq k \|x\|
\end{equation} 
holds for all $V\leq  1$ and $x \in \mathbb R^n\setminus \{0\}$ satisfying the equation $Q_1(V,x) = 0$\\
\textit{C10)} for some $k_1, k_2\in\mathbb R_+$ the inequality
\begin{equation}
   \label{strus20}
\begin{array}{cc}
k_1 \|x\|\leq V\leq k_2 \|x\|, 
\end{array}
\end{equation}
holds for all $V\geq 1$ and $x \in \mathbb R^n\setminus \{0\}$ satisfying the equation $Q_2(V,x) = 0$\\
then the origin of the system~(\ref{sys}) is globally hyperexponentially stable with degree
$r-1$.
}

\begin{proof}
Without loss of generality, let $\alpha_r\leq 1$. 
The conditions $C1)$-$C7)$  guarantee Lyapunov stability of the system. Moreover, 
according to the proof of Theorem 4 the conditions $C1)$-$C7)$ provide $V(\Phi(t, x_0))\leq V(x_0) e^{-\rho_{r, \alpha}(t)}$ that, taking into account~(\ref{rhodef}), implies
\begin{equation}
    \label{strusled}
\frac{\alpha_r}{-\ln \left(\frac{V(\Phi(t, x_0))}{V(x_0)}\right)+\alpha_r}\leq e^{-\rho_{r-1,\alpha}(t)}.    \end{equation}
Consider the following cases:

\textbf{I.} Let $V(x)\leq 1$ and $V(x_0)\geq 1$. For $\alpha_r\leq 1$ the inequality~(\ref{strus}) implies
\begin{equation}
        \label{case1}
 \begin{array}{ll}
k\|x\| &\leq \frac{\alpha_r}{1-\ln V(x)}\\
&\leq \frac{\alpha_r V(x_0)}{ V(x_0)-\ln V(x)}\\
&= \frac{\alpha_r V(x_0)}{(V(x_0)-1)-\ln V(x)+1}\\
&\leq \frac{\alpha_r V(x_0)}{\ln (V(x_0)/ V(x))+1}\\
&\leq \frac{\alpha_r V(x_0)}{\ln (V(x_0)/ V(x))+\alpha_r}.\\
\end{array}       
\end{equation}
Then, for $t\in\mathbb R: V(\Phi(t, x_0))\leq 1$, (\ref{strus}), (\ref{strusled}) and~(\ref{case1}) imply
$$
\begin{array}{ll}
\|\Phi(t,x_0)\| &\leq \frac{\alpha_r V(x_0)}{-k\ln \left(\frac{V(\Phi(t, x_0))}{V(x_0)}\right)+k\alpha_r}\\
&\leq \frac{V(x_0)}{k}e^{-\rho_{r-1,\alpha}(t)}\\
&\leq \frac{k_2}{k} \|x_0\| e^{-\rho_{r-1,\alpha}(t)}
\end{array}
$$

\textbf{II.} Let $V(x)\leq 1$ and $V(x_0)\leq 1$. The inequality~(\ref{strus}) implies
\begin{equation}
        \label{case2}
 \begin{array}{ll}
k\|x\| &\leq \frac{\alpha_r}{1-\ln V(x)}\\
&\leq \frac{\alpha_r}{1+\ln\left( \frac{V(x_0)}{V(x)} \right)}\\
&\leq \frac{\alpha_r}{\alpha_r+\ln\left( \frac{V(x_0)}{V(x)} \right)}.
\end{array}       
\end{equation}
Then~(\ref{strus}), (\ref{strusled}) and~(\ref{case2}) provide
$$
\begin{array}{ll}
  \|\Phi(t,x_0)\| &\leq \frac{\alpha_r}{-k\ln \left(\frac{V(\Phi(t, x_0))}{V(x_0)}\right)+k\alpha_r}\\
&\leq \frac{1}{k} e^{-\rho_{r-1,\alpha}(t)}.
\end{array}
$$

\textbf{III.}  Let $V(x)\geq 1$ and $V(x_0)\geq 1$. According to~(\ref{strus20}) we have
$$
\begin{array}{ll}
 \|\Phi(t,x_0)\| &\leq \frac{1}{k_1} V(\Phi(t, x_0))\\
&\leq  \frac{1}{k_1} V(x_0) e^{-\rho_{r, \alpha}(t)}\\
&\leq \frac{k_2}{k_1} \|x_0\| e^{-\rho_{r, \alpha}(t)}\\
&\leq \frac{k_2}{k_1} \|x_0\| e^{-\alpha_r \rho_{r-1, \alpha}(t)}\\
\end{array}
$$

Thus, generalizing the considered cases, the system is globally hyperexponentially stable with degree $r-1$.
\end{proof}

Applying this result to the system~(\ref{mimosys}), (\ref{mimosys2}), let us introduce the ILF functions
\begin{equation}
    \label{ILFro}
  Q_1(V,x)=x^TD\left(\varrho(V)\right)PD\left(\varrho(V)\right)x-1, \end{equation}
$$
Q_2(V,x)=\frac{1}{V^2}x^TP x-1,
$$
where $V\in\mathbb R_+$, $\varrho(V)=\ln \left(\frac{V+e-1}{V}\right)$, $D(\lambda)=\text{diag}\{\lambda^{q_i}\}_{i=1}^n$, $q_i=1+(n-i)\mu$, $\mu\in (0,1]$, $\lambda\in\mathbb R_+$ and $0<P\in\mathbb R^{n\times n}$. Denote $H=\text{diag}\{q_i\}_{i=1}^n$.
The following theorem is on rated hyperexponential control design for the linear system~(\ref{mimosys}), (\ref{mimosys2}).

\textbf{Theorem 5} \textit{Let the system of matrix inequalities
\begin{equation}
    \label{LMIHypC}
    \begin{array}{cc}
    AX + XA^T + BY + Y^T B^T + \gamma (XH+HX) \leq 0,\\
 XH + HX > 0, \quad X > 0 
 \end{array}
\end{equation}
be feasible for some $\gamma\in\mathbb R_+$ and $X\in\mathbb R^{n\times n}$, $Y\in\mathbb R^{1 \times n}$. Let 
\begin{equation}
    \label{hypcon}
  u(V,x)=\left\{\begin{array}{ll}
\varrho^{\mu-1}(V)KD(\varrho(V))x &\quad \text{for} \quad x^TPx<1,  \\
Kx &\quad \text{for} \quad x^TPx\geq 1,
  \end{array} \right.
\end{equation}
where  $K=YX^{-1}$, $X=P^{-1}\in\mathbb R^{n\times n}$, $Y\in\mathbb R^{1\times n}$ and $V\in\mathbb R_+: Q_1(V,x)=0$ for $x^TPx<1$.
Then the closed-loop system~(\ref{mimosys}), (\ref{hypcon}) is hyperexponentially stable with degree $r=1$.
}


\begin{proof}
The function $Q_1(V, x)$  satisfies the conditions \textit{C1)-C4)} of Theorem 1. Indeed, it is continuously differentiable for all $V\in\mathbb R_+$ and  $\forall x\in\mathbb R^n$. Since $P > 0$ and $\varrho(V)$ is strictly decreasing for $V\in\mathbb R_+$,  then the following chain of
inequalities 
$$
\begin{array}{ll}
\lambda_{\min}(P)\|x\|^2 \min\left\{\varrho^{2+2(n-1)\mu}(V), \varrho^2(V) \right\} \!\leq\! Q_1(V,x)\!+\!1 \\
\leq\lambda_{\max}(P)\|x\|^2 \max\left\{\varrho^{2+2(n-1)\mu}(V), \varrho^2(V) \right\}
\end{array}
$$
implies that for any $x\in\mathbb R^n \setminus\{0\}$ there exist $V^-\in\mathbb R_+$ and $V^+\in\mathbb R_+$: $Q_1(V^-,x) < 0 < Q_1(V^+, x)$.
Moreover, if $Q_1(V, x) = 0$ then the same chain of inequalities gives
\begin{equation}
    \label{normus}
 \begin{array}{ll}
\frac{1}{\lambda_{\max}(P) \max\left\{\varrho^{2+2(n-1)\mu}(V), \varrho^2(V) \right\}}  \leq \|x\|^2\leq \\
\qquad\frac{1}{\lambda_{\min}(P) \min\left\{\varrho^{2+2(n-1)\mu}(V), \varrho^2(V) \right\}} 
\end{array}   
\end{equation}
Then it follows that the conditions \textit{C2), C3)} of Theorem 1 hold.

Since
$$
\begin{array}{ll}
\frac{\partial Q_1(V,x)}{\partial V}=\!\!\!&-\frac{e-1}{V(V+e-1)\varrho(V)}\\
&\times \;x^T D\left(\varrho(V)\right) \left(PH+HP \right) D\left(\varrho(V)\right) x
\end{array}
$$
 then~(\ref{LMIHypC}) and $P:=X^{-1}$ implies $HP + PH > 0$ and $\frac{\partial Q_1}{\partial V} < 0$ for all $V\in\mathbb R_+$ and $x\in\mathbb R^n\setminus \{0\}$, and the condition \textit{C4)} of Theorem 1 also hold.
 
Now let us check that the condition \textit{C6)} holds for all $V\in(0, 1]$ and $x \in \mathbb R^n\setminus \{0\}$. Taking into account~(\ref{LMIHypC}) with $P:=X^{-1}$ and $K=YP$ we have \small
$$
\begin{array}{ll}
\frac{\partial Q_1(V, x)}{\partial x}\dot x =
2x^T D\left(\varrho(V)\right) P D\left(\varrho(V)\right)  (Ax+Bu)\\
=\!\varrho^\mu(V) x^T \!D\!\left(\varrho(V)\right)\!(\!PA\!+\!A^TP\!+\!PBK\!+\!K^TB^TP)D\!\left(\varrho(V)\right)\! x\\
\leq -\gamma \varrho^\mu(V) x^T D\left(\varrho(V)\right)(PH+HP)D\left(\varrho(V)\right) x\\
\leq \gamma V \frac{V+e-1}{e-1}\varrho^{1+\mu}(V) \frac{\partial Q_1(V,x)}{\partial V}\\
\leq \gamma V \varrho^{1+\mu}(V) \frac{\partial Q_1(V,x)}{\partial V}.\\
\end{array}
$$ \normalsize
Note that $\varrho(V)\geq -\ln(V)+\ln(e-1)$ and there exists sufficiently small $\alpha_1=\alpha_1(\mu)\in\mathbb R_+$ such that $\varrho^\mu(V)\geq \ln\left(\frac{-\ln V+\ln(e-1)}{\ln(e-1)}\right)+\alpha_1$ for $V\in(0,1]$, i.e.,
$$
\frac{\partial Q_1(V, x)}{\partial x}\dot x \leq \gamma V \sigma_1^\alpha(V) \sigma_2^\alpha(V) \frac{\partial Q_1(V,x)}{\partial V} \quad \forall V\in(0,1]
$$
with $\alpha=(\gamma, \alpha_1, \ln(e-1))^\top$.

Due to $\varrho(1)=1$ we have $Q_1(1,x) = Q_2(1,x)=x^TPx-1$, i.e., the condition $C5)$ is satisfied.

From~(\ref{normus}) for $V\leq 1$ we have
$$
\begin{array}{ll}
\|x\|&\leq \frac{1}{\sqrt{\lambda_{\min}(P)} \varrho(V)}\\ 
&\leq  \frac{2.2 \ln(e-1)}{\sqrt{\lambda_{\min}(P)} \left(-\ln V+1 \right)},
\end{array}
$$
and thus, the inequality~(\ref{strus}) holds with $k=\frac{\sqrt{\lambda_{\min}(P)}}{2.2}$.

Finally, since $V\in\mathbb R_+:Q_2(V,x)=0$ corresponds to the quadratic Lyapunov function $V(x)=(x^TPx)^{1/2}$ the further proof is straightforward. 
\end{proof}

\textbf{Remark 1} In order to calculate the function $V(x)$ implicitly defined by~(\ref{ILFro}) the bisection method may be utilized (see, e.g., \cite{nolcosmain}):

\textbf{Algorithm}

\texttt{INITIALIZATION: 
$V_0 = 1$; $a = V_{\min}$; $b = 1$;}

\texttt{STEP :}

\texttt{\quad If $x_i^T D(\varrho (b)) P D(\varrho (b)) x_i> 1$ then \\
\text{\qquad\qquad\qquad} $a = b$; $b = 2b$;}

\texttt{\quad elseif $x_i^T D(\varrho (a)) P D(\varrho (a)) x_i\!<\! 1$   then\\
\text{\qquad\qquad\qquad} $b = a$; $a = \max\left\{\frac{a}{2}, V_{\min} \right\}$;}

\texttt{\quad else\\
\text{\qquad\;\; $c=\frac{a+b}{2}$}}

\texttt{\quad\;\;\; If $x_i^T D(\varrho (c)) P D(\varrho (c)) x_i\!<\! 1$  then\\
\text{\qquad\qquad\qquad} $b = c$;}

\texttt{\quad\qquad\qquad\;\!\! else $a = \max\{V_{\min}, c\};$}

\texttt{\quad\;\;\;endif;}

\texttt{\quad endif;}

\texttt{\quad $V_i = b$;}

If \texttt{STEP} is applied recurrently many times to the same vector $x_i$ then it allows to localize  the unique positive root of the equation $x^TD\left(\varrho(V)\right)PD\left(\varrho(V)\right)x-1=0$.

\textbf{Example 2} Consider the system~(\ref{mimosys}) for $n=3$. Let us choose the same $P$, $K$ as in Example 1, and $\mu=0.2$  
that satisfy~(\ref{LMIHypC}) with $X=P^{-1}$, $Y=KX$. 
The results of simulation are shown in Fig. 4, Fig. 5. Fig. 6 demonstrates plots of the state norm $\|x\|$ for the hyperexponential control~(\ref{hypcon}) (red line) and finite-time control~(\ref{contrmimo}) (blue line). Fig. 7  demonstrates plots of  $\|x\|$ for the hyperexponential  and finite-time controls  with measurement band limited noise of power $10^{-5}$.
Fig. 8  shows plots of the state norm $\|x\|$ in the presence of delay $\tau=0.05$ in the control channel. It is easy to see that the control~(\ref{hypcon}) has better robustness properties than the finite-time control~(\ref{contrmimo}).  A detailed study of the presented control algorithms on robustness analysis with
respect to disturbances, uncertainties, delays and extension of these results on a wider class of systems goes beyond the scope of the paper providing the subjects for a future research.

 \begin{figure}
\begin{center}
\includegraphics[width=8.5cm]{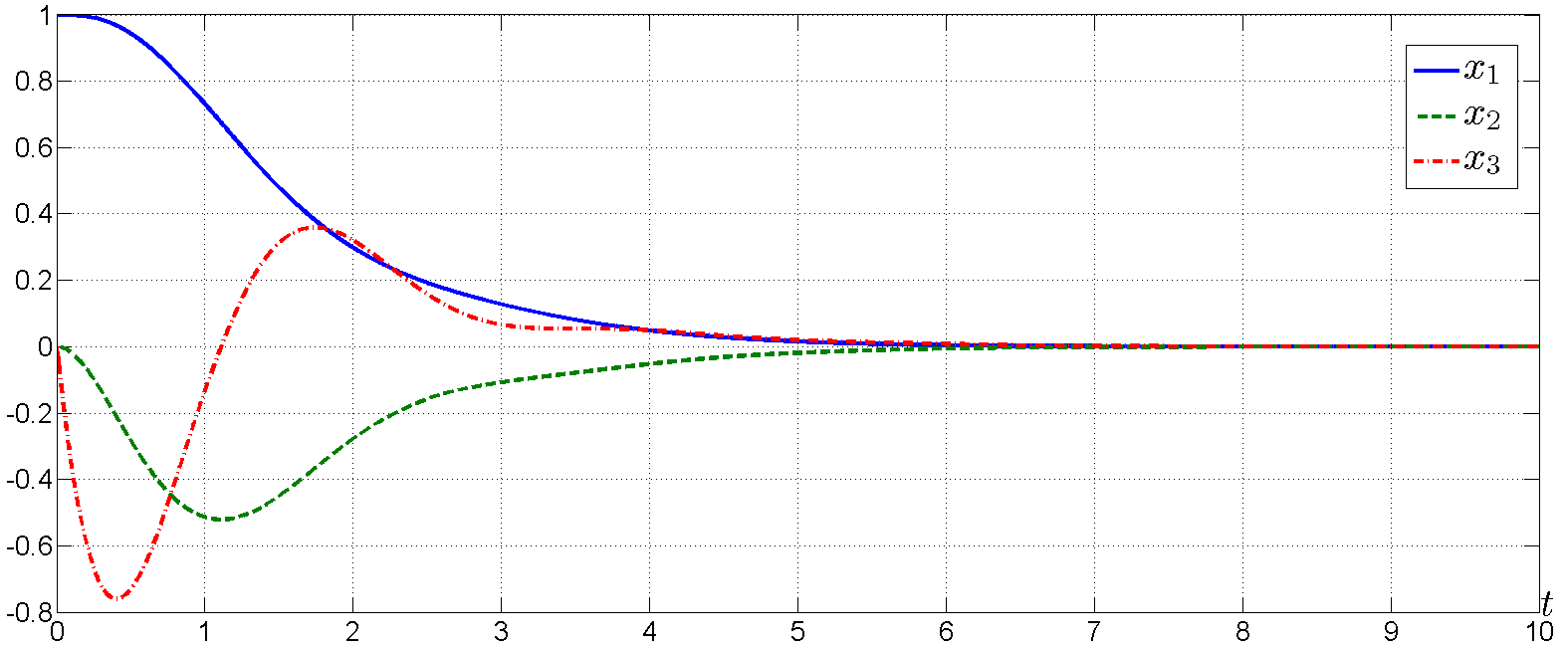}   
\caption{State transients for the system~(\ref{mimosys}), (\ref{hypcon}) }
\label{Fig4}                              
\end{center}                                
\end{figure}

 \begin{figure}
\begin{center}
\includegraphics[width=8.5cm]{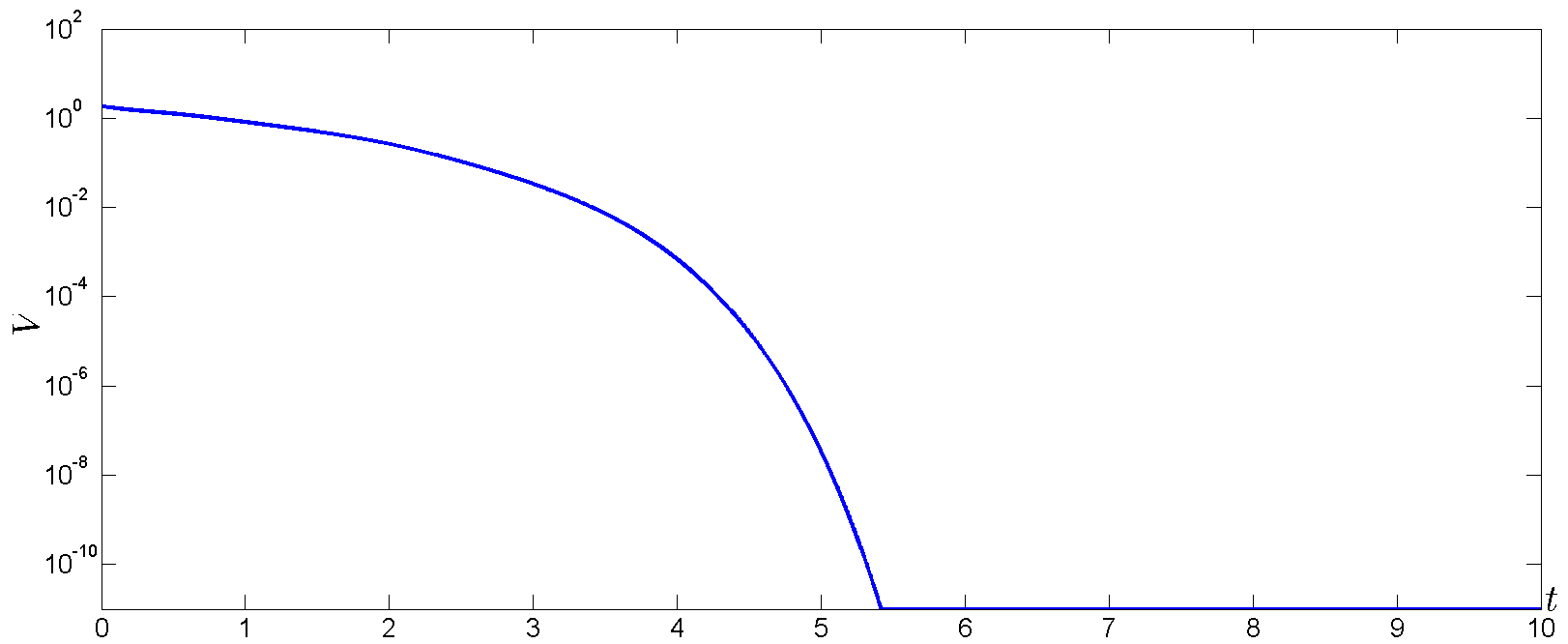}   
\caption{Lyapunov function for the system~(\ref{mimosys}), (\ref{hypcon}) }
\label{Fig5}                              
\end{center}                                
\end{figure}

 \begin{figure}
\begin{center}
\includegraphics[width=8.5cm]{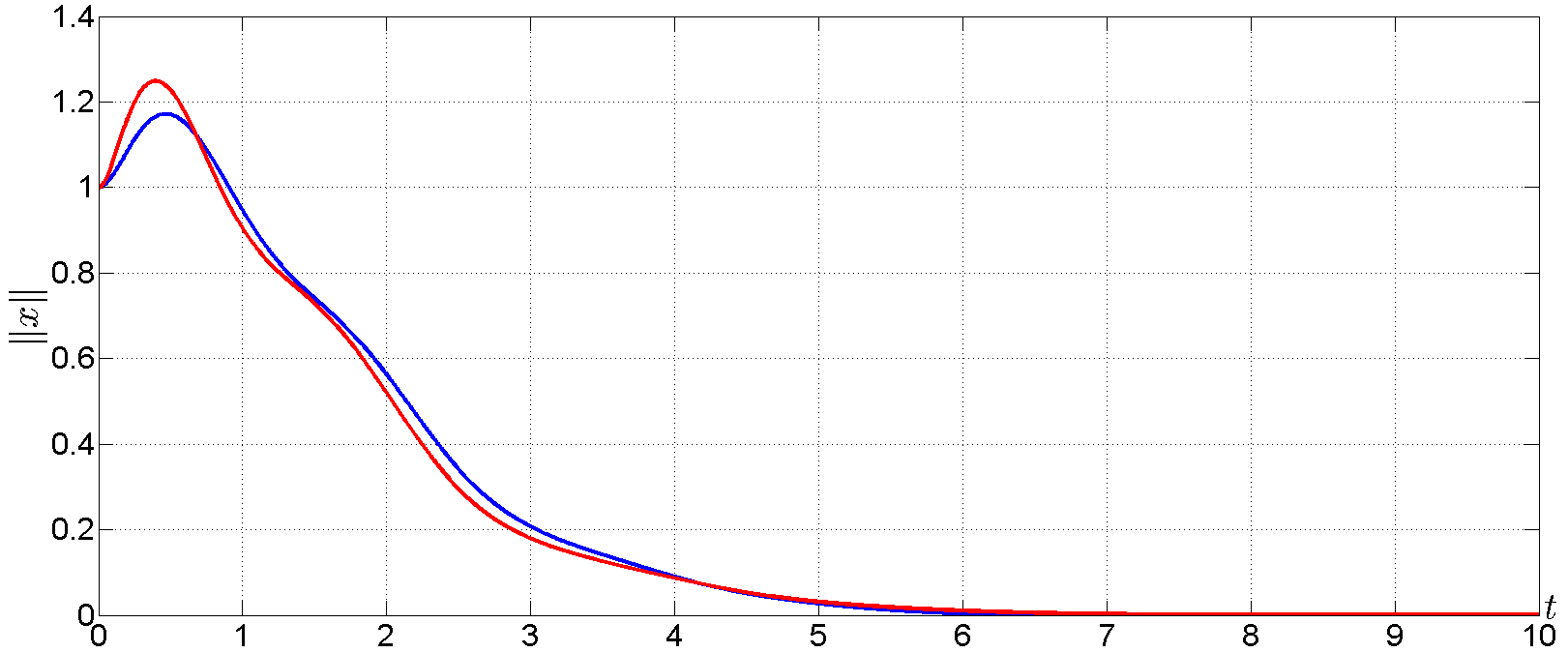}   
\caption{Simulation plot of $\|x\|$ for the hyperexponential (red line) and finite-time (blue line) control with initial conditions $x(0)=[1\; 0\; 0]^T$), (\ref{hypcon}) }
\label{Fig6n}                              
\end{center}                                
\end{figure}

 \begin{figure}
\begin{center}
\includegraphics[width=8.5cm]{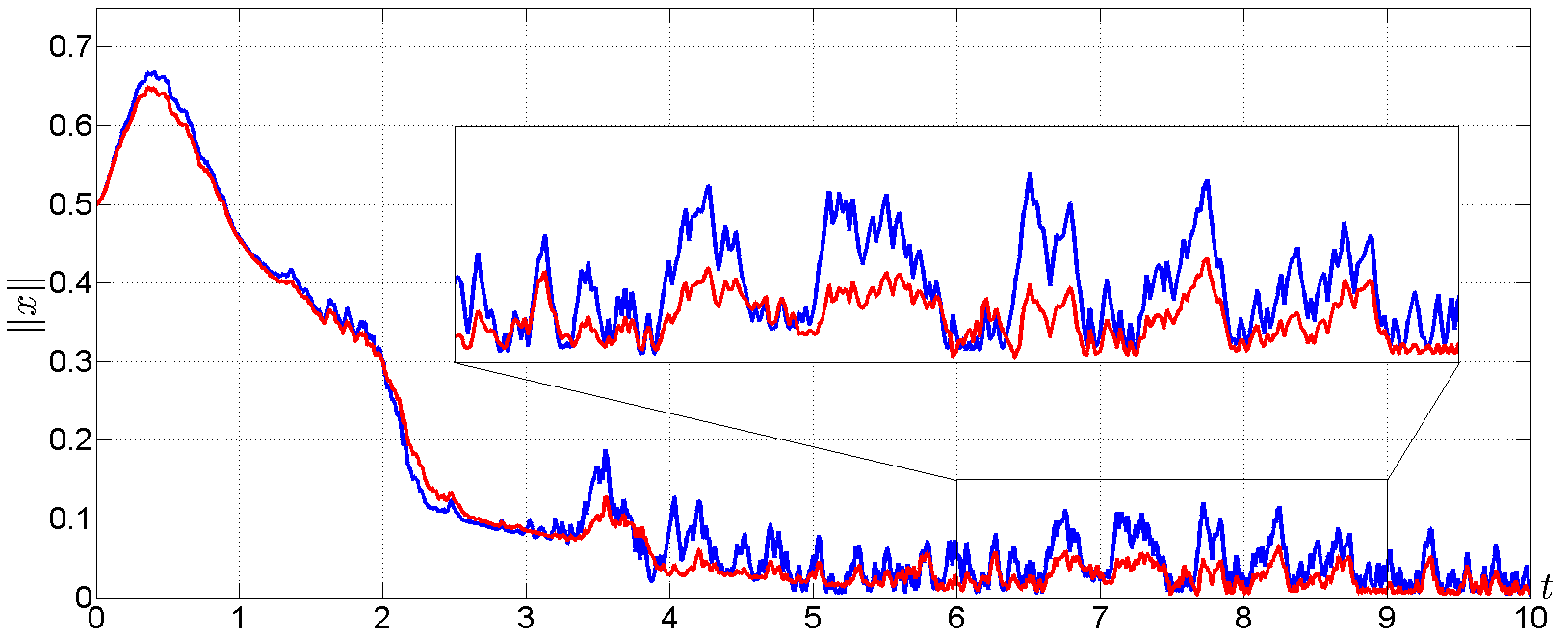}   
\caption{Simulation plot of $\|x\|$ for the hyperexponential (red line) and finite-time (blue line) control with measurement band limited noise of power $10^{-5}$
and initial conditions $x(0)=[0.5\; 0\; 0]^T$
}
\label{Fig6}                              
\end{center}                                
\end{figure}

 \begin{figure}
\begin{center}
\includegraphics[width=8.5cm]{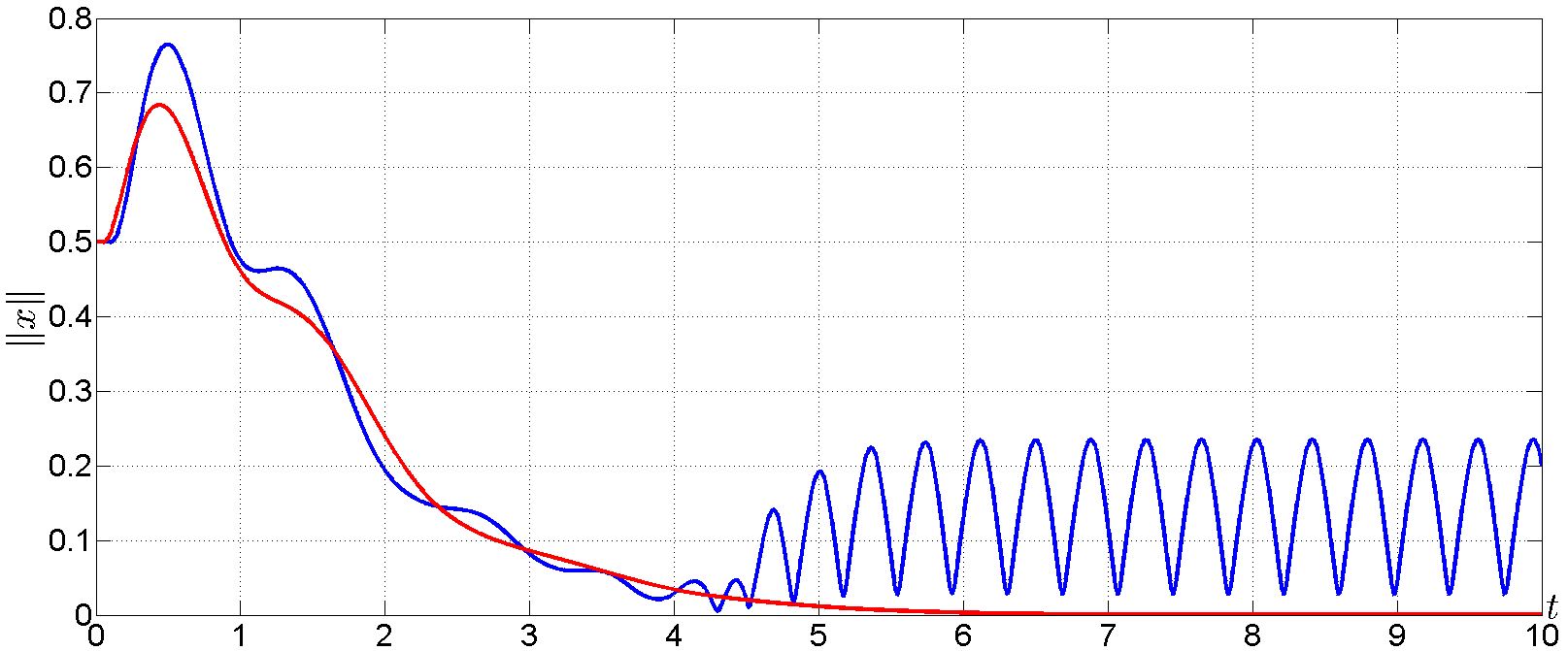}   
\caption{Simulation plot of $\|x\|$ for the hyperexponential (red line) and finite-time (blue line) control in the presence of delay $\tau=0.05$ in the control channel for initial conditions $x(0)=[0.5\; 0\; 0]^T$}
\label{Fig6}                              
\end{center}                                
\end{figure}


In order to ensure fast convergence outside the vicinity of the origin one can combine a nearly fixed-time control with hyperexponential one.

Define 
$$ 
\bar Q_2(V,x)=x^T \bar D\left(V^{-1}\right)P\bar D\left(V^{-1}\right)x-1,  
$$
where $\bar D(\lambda)=\text{diag}\{\lambda^{p_i}\}_{i=1}^n$, $p_i=1+(i-1)\nu$, $\nu\in\mathbb R_+$, $\lambda\in\mathbb R_+$ and $0<P\in\mathbb R^{n\times n}$. Denote $\bar H=\text{diag}\{p_i\}_{i=1}^n$.

\pagebreak
\textbf{Corollary 1} \textit{Let the system of matrix inequalities~(\ref{LMIHypC}) and
$$
 X\bar H +\bar HX > 0
$$
be feasible 
for some $\gamma\in\mathbb R_+$ and $X\in\mathbb R^{n\times n}$, $Y\in\mathbb R^{1 \times n}$. Let 
\begin{equation}
    \label{hypcon2}
  u(V,x)=\left\{\begin{array}{ll}
KD(\varrho(V))x &\quad \text{for} \quad x^TPx<1,  \\
V^{1+\nu} K\bar D(V^{-1}) x &\quad \text{for} \quad x^TPx\geq 1,
  \end{array} \right.
\end{equation}
where  $K=YX^{-1}$, $X=P^{-1}\in\mathbb R^{n\times n}$, $Y\in\mathbb R^{1\times n}$ and 
 $$
 V\in\mathbb R_+:\left\{\begin{array}{ll}
  Q_1(V,x)=0 \quad \text{for} \quad x^TPx<1,\\
\bar   Q_2(V,x)=0 \quad \text{for} \quad x^TPx\geq 1.\\
  \end{array}
 \right.
 $$ 
Then the closed-loop system~(\ref{mimosys}), (\ref{hypcon2}) is globally nearly  fixed-time and  hyperexponentially stable with degree $1$.}

\begin{proof}
The proof follows exactly the same arguments as one of Theorem 5 and \cite[Theorem 9]{ILF_Automatica}.
\end{proof}


Using the same arguments as in Proposition 1 it is easy to show that under sampled-time realization the ILF-based  control~(\ref{hypcon}) preserves hyperexponential stability: 

\textbf{Corollary 2} \textit{Let $\{t_i\}_{i=0}^\infty$ be a strictly increasing sequence of arbitrary time instants, $0 = t_0 < t_1 < t_2 <...$ such that $\lim_{i\to\infty} t_i = +\infty$. Let for~(\ref{mimosys}) all conditions of Theorem 5 (Corollary 1)  be satisfied. Then sampled-time realization $u(t) = u(V_i, x(t))$ for $t\in [t_i, t_{i+1})$, where $V_i\in\mathbb R_+ : Q(V_i, x(t_i)) = 0$ provides hyperexponential stability of the closed-loop system.}


\section{CONCLUSIONS}

In the paper sufficient Lyapunov characterizations
of hyperexponential stability are presented for the
system~(\ref{sys}) using explicit and implicit Lyapunov function approaches. Firstly, it is shown that ILF-based finite/fixed-time control methods provide hyperexponential stability under sampled-time realization. Next, the hyperexponential control~(\ref{hypcon}) was proposed for the linear system~(\ref{mimosys}). The preliminary numeric experiments indicate that this control is  
less sensitive with respect to noises than  its finite-time analog. In addition, the hyperexponential control demonstrates better performance in the presence of delays as well.
Tuning control parameters is presented
in the form of linear matrix  inequalities. The performance of the proposed control  is illustrated through simulations.

The presented results open a lot of topics for future research. For example, development of ILF-based hyperexponential controls and observers for nonlinear and MIMO systems, detailed study of the presented control on robustness analysis with
respect to disturbances, uncertainties, delays, etc.

\addtolength{\textheight}{-13.5cm}   



\begin{thebibliography}{99}

\bibitem{Caraballo} T. Caraballo, On the Decay Rate of Solutions of Non-Autonomous Differential Systems, Electronic Journal of Differential Equations, vol. 2001(05), pp. 1--17, 2001.


\bibitem{Palamarchuk} E. Palamarchuk, On the Generalization of Logarithmic Upper Function for Solution of a Linear Stochastic Differential Equation with a Nonexponentially Stable Matrix, Differential Equations, vol. 54(2), pp. 193--200, 2018.

\bibitem{Filippov} A. Filippov, Differential equations with discontinuous right-hand sides, Dordrecht, Kluwer, 1988.

  \bibitem{bernstein} S. Bhat, D. Bernstein, Finite-time stability of continuous autonomous systems, SIAM Journal of Control and Optimization, vol. 38(3), pp. 751-766, 2000.
  
  \bibitem{Orlov2004} Y. Orlov, Finite Time Stability and Robust Control Synthesis
of Uncertain Switched Systems,  SIAM Journal of Control and Optimization, vol. 43(4), pp. 1253-1271, 2004.

\bibitem{fixattr} A. Polyakov, Nonlinear Feedback Design for Fixed-Time Stabilization of Linear Control Systems, IEEE Transactions on Automatic Control, vol. 57(8), pp. 2106–2110, 2012.

\bibitem{Polyakov_ILF_TD} A. Polyakov, D. Efimov, W. Perruquetti, J.-P. Richard, 
Implicit Lyapunov-Krasovski Functionals for Stability Analysis and Control Design of Time-Delay Systems, IEEE Transactions on Automatic Control, vol. 60(12), pp. 3344–3349, 2015.


\bibitem{Polyakov_Fast_C} A. Polyakov, Fast Control Systems: Nonlinear Approach, In: Clempner J., Yu W. (eds) New Perspectives and Applications of Modern Control Theory, Springer, Cham, 2018. https://doi.org/10.1007/978-3-319-62464-8\_12


\bibitem{Lyapunov_1992} A.M. Lyapunov, The general problem of the stability of motion, Taylor \& Francis, 1992.

\bibitem{Korobov} V.I. Korobov, A general approach to synthesis problem,
Doklady Academii Nauk SSSR, vol. 248, pp. 1051--1063, 1979.

\bibitem{Adamy_Flemming} J. Adamy, A. Flemming, Soft variable-structure
controls: a survey, Automatica, vol. 40, pp. 1821--1844, 2004.

\bibitem{Clarke_1990} F. Clarke, Optimization and Nonsmooth Analysis, SIAM, Philadelphia, 1990.

\bibitem{Efimov_FT_delay} D. Efimov, A. Polyakov, E. Fridman, W. Perruquetti, J.-P. Richard, Comments on finite-time stability of time-delay systems, Automatica, vol. 50(7), pp. 1944--1947, 2014.


\bibitem{Nekhoroshikh2020} A. Nekhoroshikh, D. Efimov, A. Polyakov, W. Perruquetti, I. Furtat, On Finite-Time Stabilization of a Class of Nonlinear Time-Delay Systems: Implicit Lyapunov-Razumikhin Approach, Proc. 59th IEEE Conference on Decision and Control (CDC), Republic of Korea, 2020.




\bibitem{Shen_2020} W. Shen, X. Wang, H. Liu, X. Zhang, B. Cai, Global Exponential Stability Criteria for Proportional Delay High-Order Neural
Networks: A Hyper-Exponential Stability Technique, Preprints of the 21st IFAC World Congress, Berlin, Germany, 2020.

\bibitem{ILF_Automatica} A. Polyakov, D. Efimov, W. Perruquetti, Finite-time and fixed-time stabilization: Implicit Lyapunov function approach, Automatica, vol. 51, pp. 332--340, 2015.

\bibitem{mimo} A. Polyakov, D. Efimov, W. Perruquetti, Robust stabilization of MIMO systems in finite/fixed time, International Journal of Robust and Nonlinear Control, vol. 26(1), pp. 69--90, 2016.

\bibitem{Zimenko_TAC} K. Zimenko, A. Polyakov, D. Efimov, W. Perruquetti, Robust Feedback Stabilization of Linear MIMO Systems Using Generalized Homogenization, IEEE Transactions on Automatic Control, vol. 65(12), pp. 5429--5436, 2020.

\bibitem{Efimov_DI_2018} D. Efimov, E. Fridman, W. Perruquetti, J.-P. Richard,  On hyper-exponential output-feedback stabilization of a double integrator by using artificial delay, 2018 European Control Conference,  pp. 1178--1182, 2018.

\bibitem{nolcosmain} A. Polyakov, D. Efimov, W. Perruquetti, Finite-time stabilization using implicit lyapunov function technique, IFAC Proceedings Volumes, vol. 46(23), pp. 140-145, 2013.

\bibitem{Polyakov_CDC_17} A. Polyakov, D. Efimov, E. Fridman, W. Perruquetti, J.-P. Richard, On hyper exponential stabilization of linear state-delay systems,
56th Annual Conference on Decision and Control (CDC), pp. 4321--4326, 2018.


\end{thebibliography}
\end{document}